\documentclass[aps,prl,superscriptaddress,reprint,floatfix,longbibliography]{revtex4-1}
\usepackage[version=3]{mhchem} 
\usepackage{units}
\usepackage{graphicx}
\usepackage{color}
\usepackage{subfigure}
\usepackage{multirow}
\usepackage{array}

\graphicspath{ {./Figures/} }

\newcommand{\wn}{cm$^{-1}$}

\newcommand{\Etwog}{\textrm{E}_{2\textrm{g}}}
\newcommand{\Aoneprime}{A'_{1}}
\newcommand{\Atwoprime}{A'_{2}}

\newcommand{\etal}{\emph{et al.}}
\newcommand{\iph}{I-Ph}

\begin{document}

\bibliographystyle{naturemag}

\title{Phonon-Induced Transparency in Functionalized Single Layer Graphene}

\author{Bruno Rousseau} 
\affiliation{
    Regroupement Qu\'{e}b\'{e}cois sur les Mat\'{e}riaux de Pointe  and
    D\'{e}partement de physique, Universit\'{e} de Montr\'{e}al, C.~P. 6128, 
    Succursale Centre-Ville, Montr\'{e}al, Qu\'{e}bec H3C~3J7, Canada}

\author{Fran\c{c}ois Lapointe}
\affiliation{
    Regroupement Qu\'{e}b\'{e}cois sur les Mat\'{e}riaux de Pointe  and
    D\'{e}partement de chimie, Universit\'{e} de Montr\'{e}al,
    C.~P. 6128, Succursale Centre-Ville, Montr\'{e}al, Qu\'{e}bec H3C~3J7, Canada}

\author{Minh Nguyen}
\affiliation{
    Regroupement Qu\'{e}b\'{e}cois sur les Mat\'{e}riaux de Pointe  and
    D\'{e}partement de chimie, Universit\'{e} de Montr\'{e}al,
    C.~P. 6128, Succursale Centre-Ville, Montr\'{e}al, Qu\'{e}bec H3C~3J7, Canada}

\author{Maxime Biron} 
\affiliation{
    Regroupement Qu\'{e}b\'{e}cois sur les Mat\'{e}riaux de Pointe  and
    D\'{e}partement de g\'{e}nie physique, \'{E}cole Polytechnique de Montr\'{e}al, C.~P. 6079,
    Succursale Centre-ville, Montr\'{e}al, Qu\'{e}bec H3C~3A7, Canada} 

\author{Etienne Gaufr\`{e}s}
\affiliation{
    Regroupement Qu\'{e}b\'{e}cois sur les Mat\'{e}riaux de Pointe  and
    D\'{e}partement de chimie, Universit\'{e} de Montr\'{e}al,
    C.~P. 6128, Succursale Centre-Ville, Montr\'{e}al, Qu\'{e}bec H3C~3J7, Canada}

\author{Saman Choubak}
\affiliation{
    Regroupement Qu\'{e}b\'{e}cois sur les Mat\'{e}riaux de Pointe  and
    D\'{e}partement de g\'{e}nie physique, \'{E}cole Polytechnique de Montr\'{e}al, C.~P. 6079,
    Succursale Centre-ville, Montr\'{e}al, Qu\'{e}bec H3C~3A7, Canada} 

\author{Zheng Han}
\affiliation{Universit\'{e} de Grenoble, Alpes, Institut N\'{E}EL, F-38042
Grenoble, France}
\affiliation{CNRS, Institut N\'{E}EL, F-38042 Grenoble, France}

\author{Vincent Bouchiat}
\affiliation{Universit\'{e} de Grenoble, Alpes, Institut N\'{E}EL, F-38042
Grenoble, France}
\affiliation{CNRS, Institut N\'{E}EL, F-38042 Grenoble, France}

\author{Patrick Desjardins} 
\affiliation{
    Regroupement Qu\'{e}b\'{e}cois sur les Mat\'{e}riaux de Pointe  and
    D\'{e}partement de g\'{e}nie physique, \'{E}cole Polytechnique de Montr\'{e}al, C.~P. 6079,
    Succursale Centre-ville, Montr\'{e}al, Qu\'{e}bec H3C~3A7, Canada} 

\author{Michel C\^{o}t\'{e}} \email{Michel.Cote@umontreal.ca}
\affiliation{
    Regroupement Qu\'{e}b\'{e}cois sur les Mat\'{e}riaux de Pointe  and
    D\'{e}partement de physique, Universit\'{e} de Montr\'{e}al, C.~P. 6128, 
    Succursale Centre-Ville, Montr\'{e}al, Qu\'{e}bec H3C~3J7, Canada} 

\author{Richard Martel} \email{r.martel@umontreal.ca}
\affiliation{
    Regroupement Qu\'{e}b\'{e}cois sur les Mat\'{e}riaux de Pointe  and
    D\'{e}partement de chimie, Universit\'{e} de Montr\'{e}al,
    C.~P. 6128, Succursale Centre-Ville, Montr\'{e}al, Qu\'{e}bec H3C~3J7, Canada}

\begin{abstract}
    Herein, intervalley scattering is exploited to account for anomalous antiresonances in
    the infrared spectra of doped and disordered single layer graphene. 
    We present infrared spectroscopy measurements of graphene grafted with
    iodophenyl moieties in both reflection microscopy and transmission
    configurations. 
    Asymmetric transparency windows at energies corresponding to
    phonon modes near the $\Gamma$ and K points are observed, in contrast to the featureless
    spectrum of pristine graphene. 
    These asymmetric antiresonances are demonstrated to vary as a function of the
    chemical potential. 
    We propose a model which involves coherent intraband scattering with
    defects and phonons, thus relaxing the optical selection rule
    forbidding access to ${\bf q} \neq \Gamma$ phonons. 
    This interpretation of the new phenomenon is supported by our numerical
    simulations that reproduce the experimental features.
\end{abstract}

\maketitle

Graphene has found promising applications in plasmonics for the terahertz
(THz) to mid-infrared (MIR) regime because of the optical properties
pertaining to its bidimensional semimetallic nature~\cite{Low2014}.
In this energy range, its absorption spectrum 
is dominated by a strong Drude-like response that can be modulated by
doping~\cite{Horng2011}.
The optical response is, however, monotonous 
because there is no infrared-active phonon mode available, a fact explained
by Zallen's rule stating that at least three atoms in the primitive unit
cell are necessary (and sufficient) to generate infrared activity in an
elemental crystal~\cite{Zallen1968}.
Tailoring the optical response of single layer graphene (SLG) has therefore
required creating plasmonic resonators and metamaterials by patterning the
material into nanoribbons~\cite{Ju2011a,Yan2013,Brar2015}, nanodisks,
heterostructure stacks~\cite{Yan2012a}, etc.
In contrast, pristine bilayer graphene (BLG) demonstrates a native tunable
Fano resonance~\cite{Kuzmenko2009,Tang2010}, while BLG nanoribbons exhibit phonon-induced
transparency through plasmon-phonon coupling~\cite{Yan2014}.


We report on a novel scattering phenomenon mediated by phonons and
disorder, and its manifestation in the MIR spectra of single layer graphene
(SLG).  The infrared spectra of covalently functionalized graphene show
Fano-like antiresonances which we demonstrate to emerge from scattering
with randomly distributed grafts, thereby allowing momentum transfer
between phonons and electronic intraband excitations.  
The scattering process leads to sharp transparency windows in the
mid-infrared optical conductivity at frequencies corresponding to optical
phonon energies for momenta near the $\Gamma$ and $\pm$~K points, the
latter being a direct consequence of intervalley scattering. 
The mechanism is reminiscent, but distinct, of the double resonance
scattering process between the two Dirac cones of graphene, that leads to
the occurrence of the $D$~band in Raman spectroscopy of disordered
samples~\cite{Thomsen2000}. 
The phenomenon appears ubiquitous to low dimension
carbon structures such as carbon nanotubes~\cite{Lapointe2012}, and can
potentially be exploited in optoelectronic applications in the infrared.

\section*{Results}

\begin{figure*}
    \includegraphics{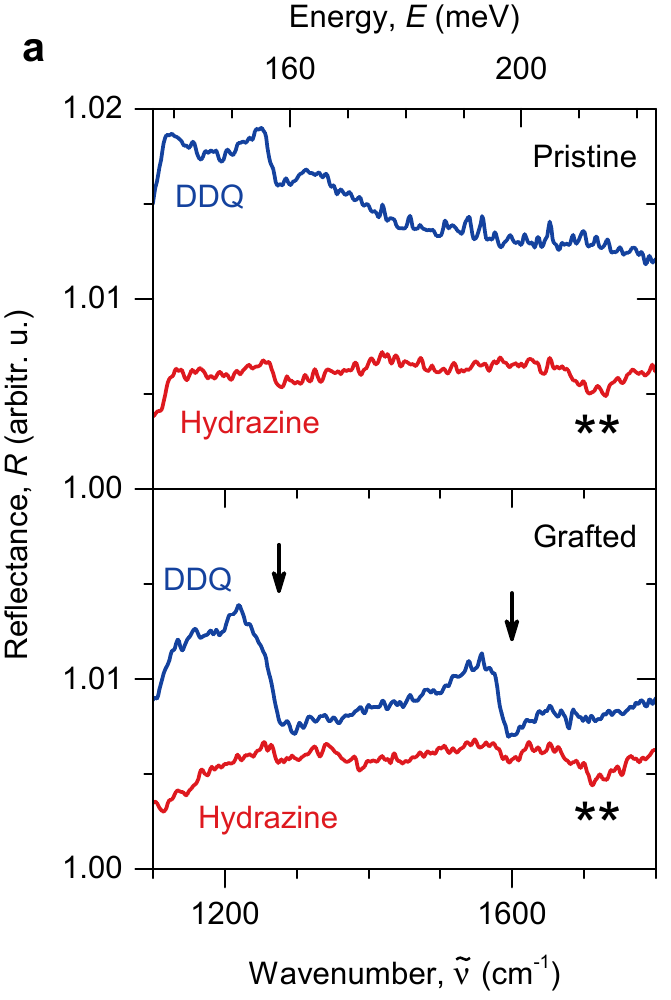}
    $\qquad$
    \includegraphics{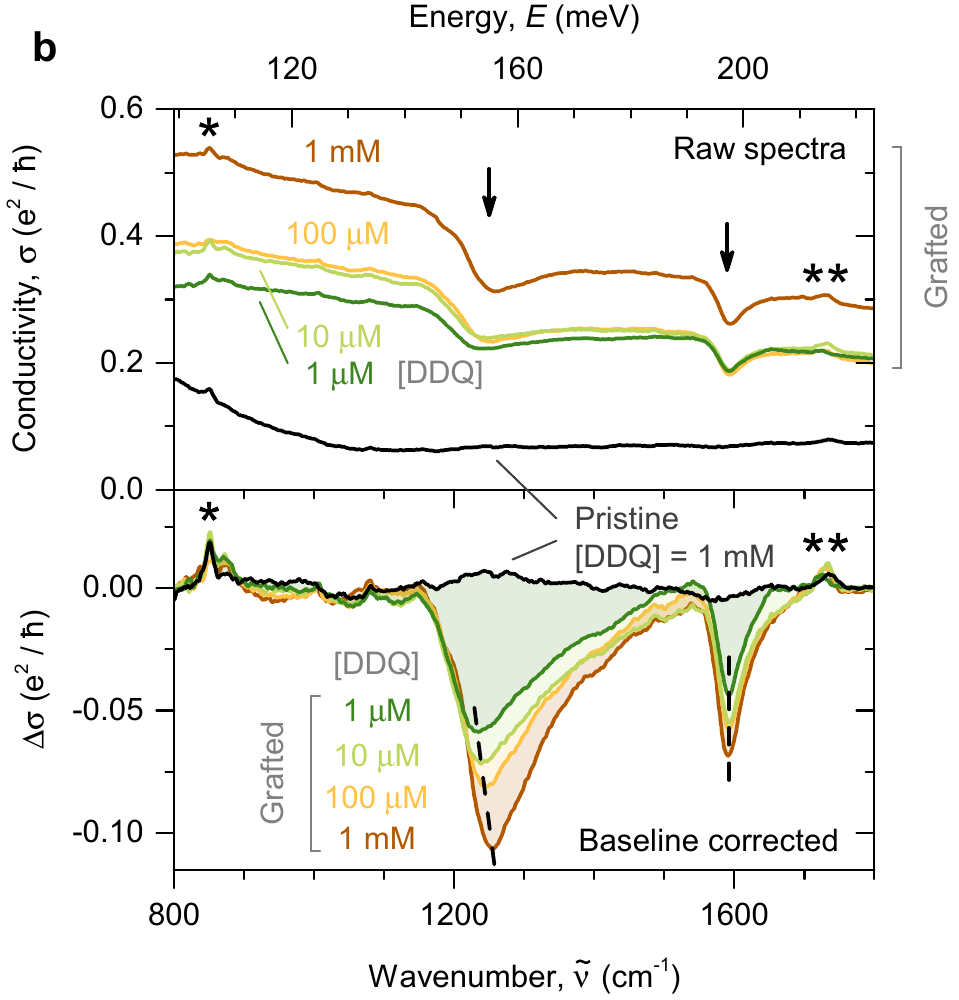}
    \caption{
        (a)~Mid-infrared reflection spectra in microscopy of pristine (upper
    panel) and electrografted single layer graphene (lower panel)
    transferred onto intrinsic silicon with 300~nm thermal oxide.
        Spectra are shown after soaking in dopant solutions of [DDQ]~$=
        1$~mM and [hydrazine]~$= 10$~mM in acetonitrile. 
    (b)~Mid-infrared optical conductivity of single layer graphene on
    BaF$_2$ measured in transmission.
    Raw spectra are presented in the upper panel, while the lower panel
    shows traces after subtraction of a cubic baseline.
    The black trace comes from a pristine sample doped at [DDQ]~$= 1$~mM.
    The other traces belong to a functionalized sample for varying DDQ
    concentrations. 
    The arrows show the disorder-induced antiresonances, while the star
    ($\ast$) denotes a spurious band, and the double stars ($\ast\ast$), the 
    carbonyl band of an impurity.
    }
    \label{fig:slg-data}
\end{figure*}

\begin{figure}
\centering
\begin{tabular}{cc}
       \setcounter{subfigure}{0}
       \subfigure[Electronic excitation with phonon scattering]{
        \includegraphics[width=0.45\linewidth]{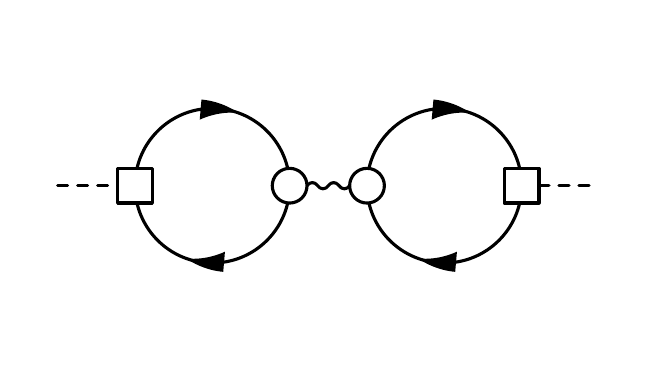}
       \label{fig:current current fano}
        } 
     &
       \setcounter{subfigure}{1}
    \subfigure[Defect scattering included]{
          \includegraphics[width=0.45\linewidth]{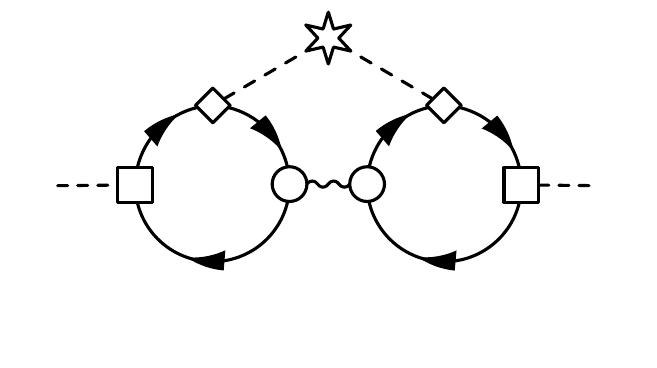}
      \label{fig:current current impurity}
                      } 
    \\
       \setcounter{subfigure}{2}
       \subfigure[Raman amplitude $\mathcal{M}$]{
        \includegraphics[width=0.45\linewidth]{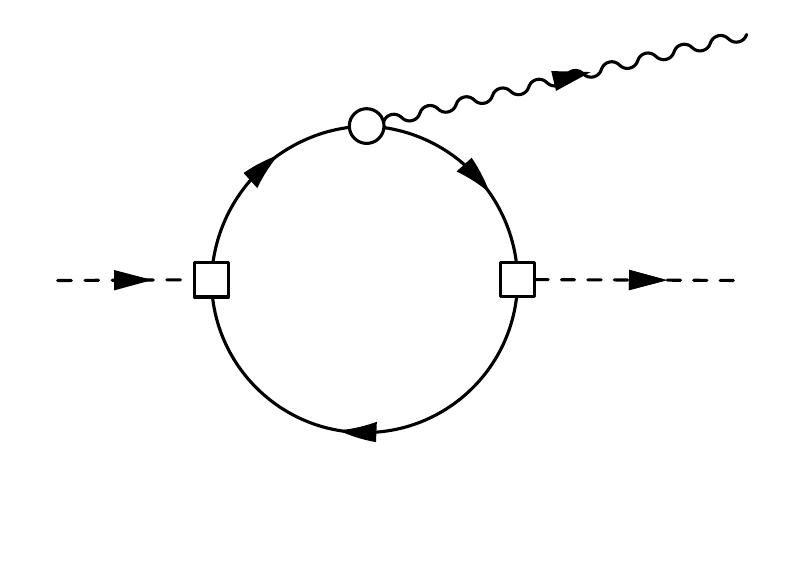}
       \label{fig:Raman}
        } 
     &
       \setcounter{subfigure}{3}
       \subfigure[Loop function $\mathcal{H}$]{
          \includegraphics[width=0.45\linewidth]{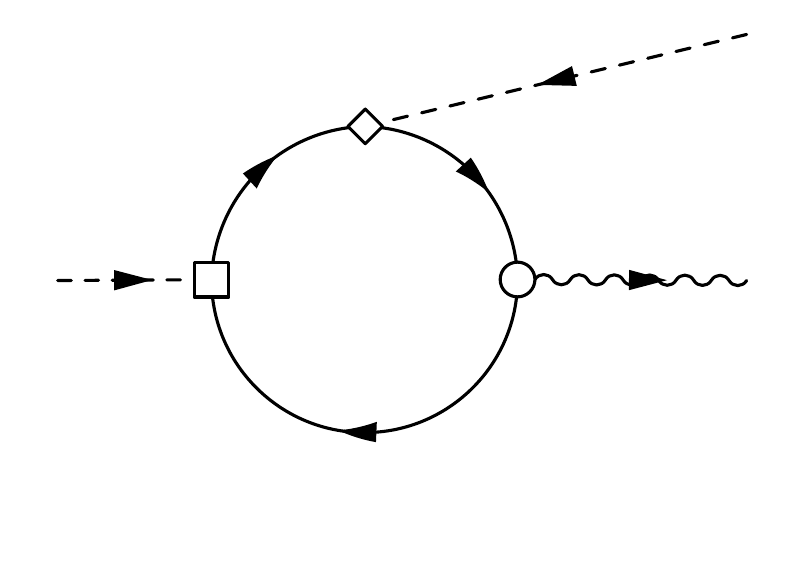}
      \label{fig:Hloop}
                      } 
    \\
\multicolumn{2}{c}{
       \setcounter{subfigure}{4}
       \subfigure[Resonance condition on $\mathcal{M}$ and  $\mathcal{H}$]{
        \includegraphics[width=0.95\linewidth]{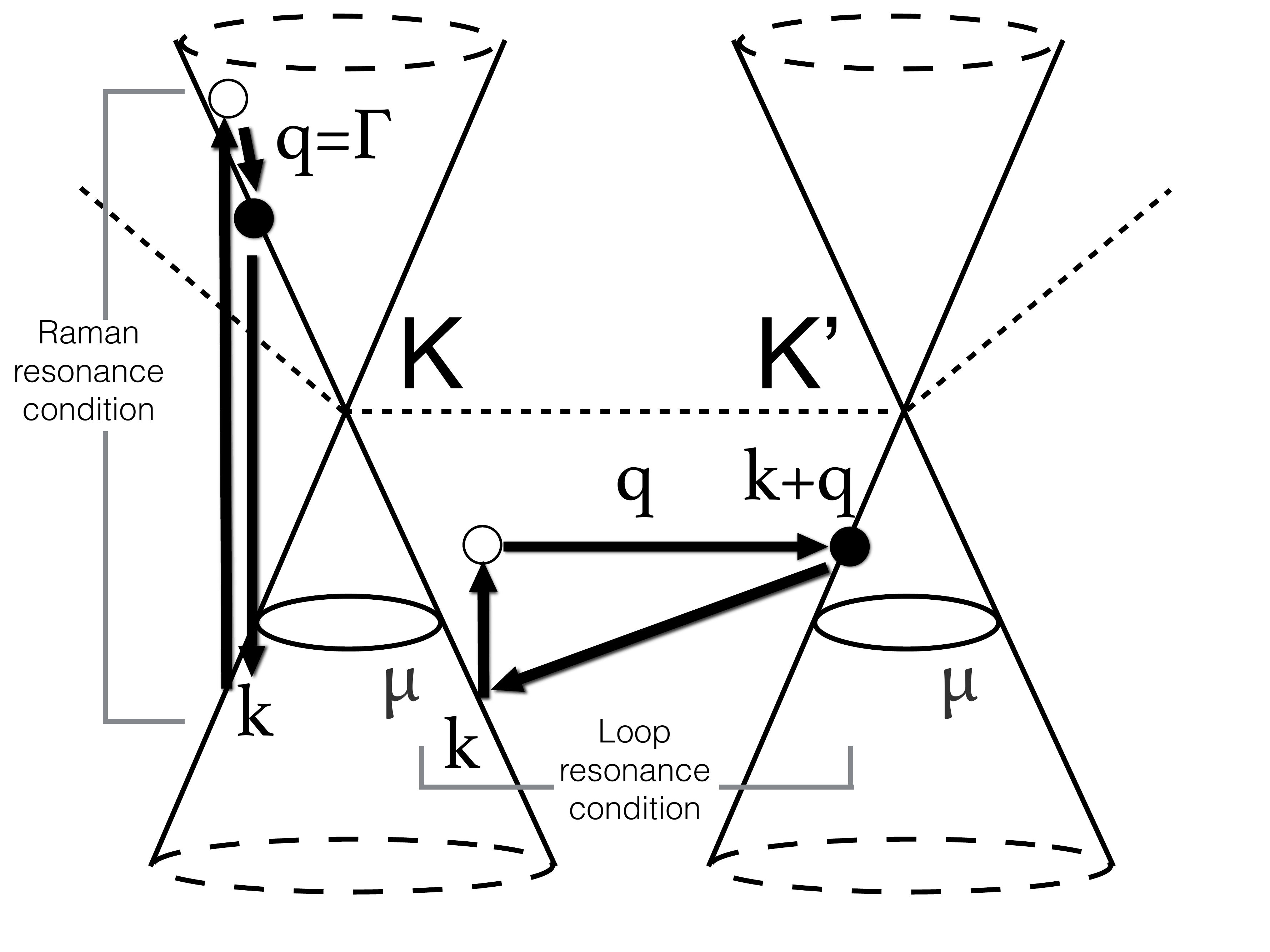}
       \label{fig:resonance condition}
        } 
}
\end{tabular}
\caption{
    (a-b-c-d) Various Feynman diagrams.
    The squares represent the matrix elements of the light-electron interaction,
    the circles represent the electron-phonon coupling,
    the diamonds the self-consistent, energy dependent defect scattering potential,
    the star corresponds to a factor of $n_{imp}$ (the number density of defects),
    the full lines the electronic Green functions
    and the wavy lines the phonon propagator.
    (a)
    Contribution to the current-current correlation
    function, which leads to a Fano profile in bilayer graphene but vanishes identically
    in single layer graphene, (b)
    extension involving both phonon and defect scattering beyond electronic
    self-energy effects, thus leading to optical bands that are forbidden
    in pristine samples.
    (c) Feynman diagram corresponding to a Raman scattering resonance.
    (d) Feynman diagram corresponding to the loop function.
    (e) Schematics of the resonant scattering mechanisms for both Raman spectroscopy and the loop function.
}
\end{figure}


\begin{figure*}
\centering
\begin{tabular}{cc}
    \subfigure[DOS and conductivity]{\includegraphics[width=0.5\linewidth]{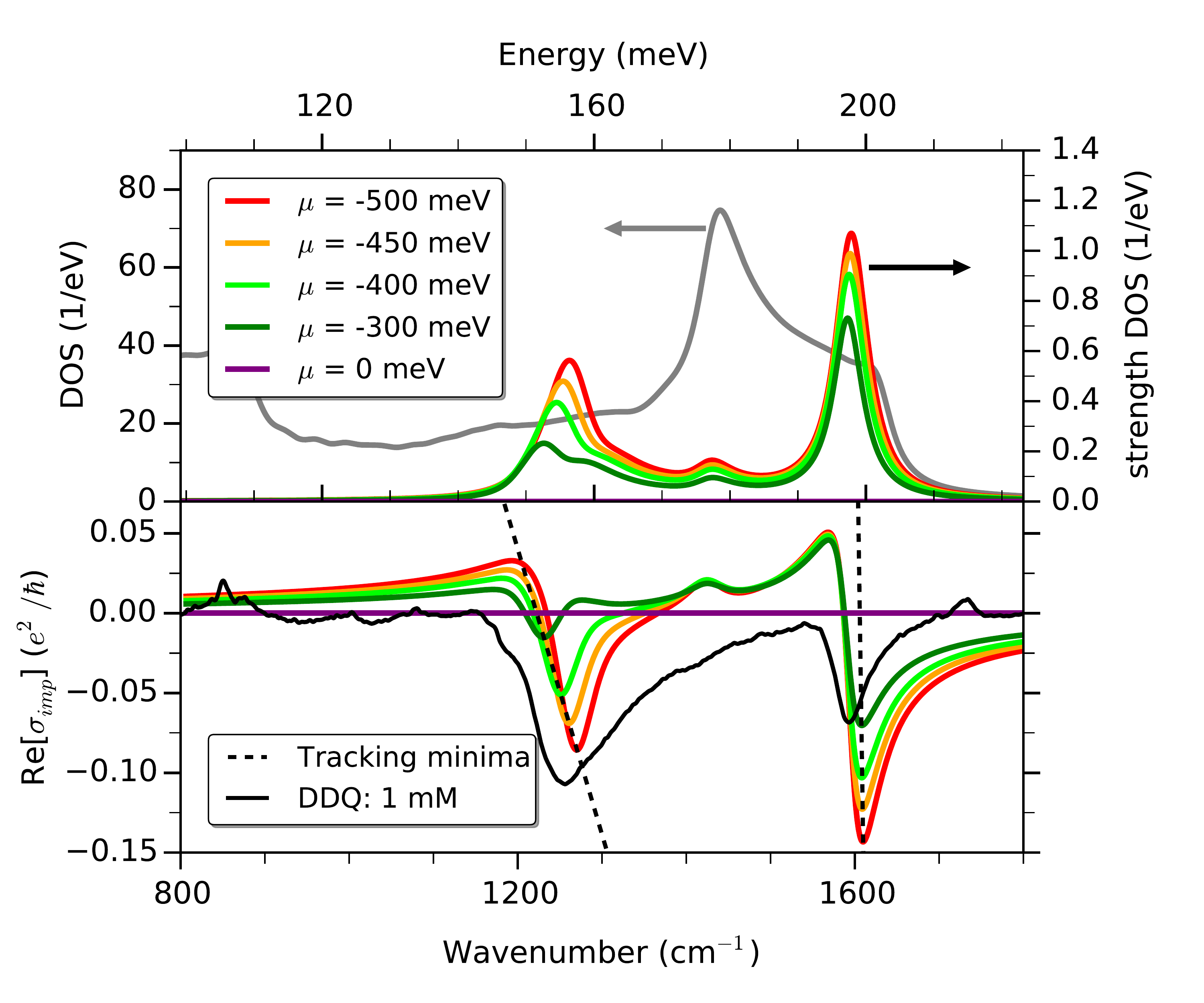} 
    \label{fig:phonon strength} }  
    &
    \subfigure[Phonon dispersion and coupling strength]{ \includegraphics[width=0.5\linewidth]{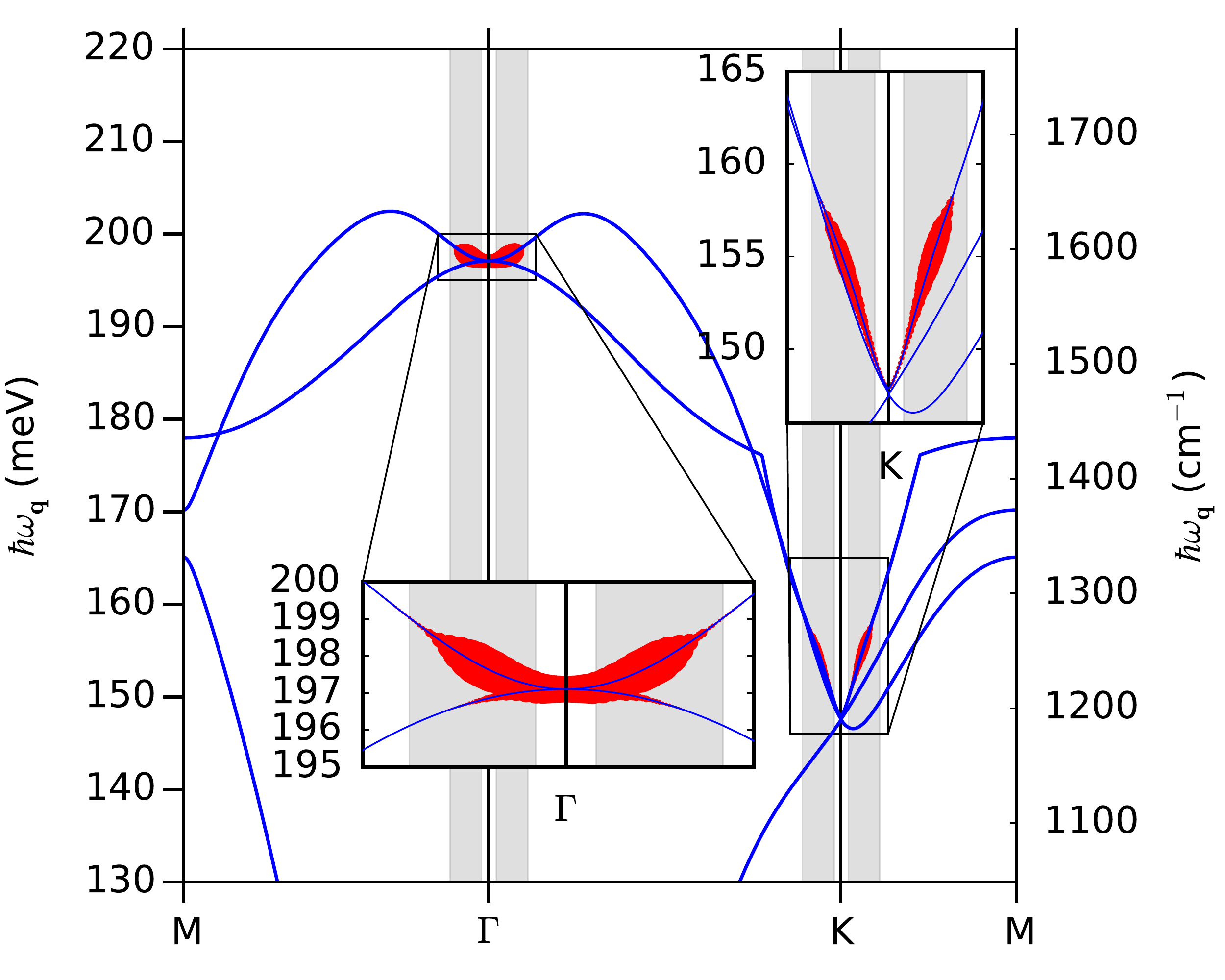}
     \label{fig:phonon dispersion} }
    \\
\end{tabular}
\caption{\label{fig:results}
    (a)
     Computed phonon density of states and density of states weighted with the phonon coupling strength $P_{\nu \bf q}$ (top panel);
     computed contributions to the conductivity for various chemical
     potential values, compared to experimental 
     results corresponding to [DDQ]~$= 1$~mM (lower panel).
    (b)
    Model phonon dispersion of graphene (thin blue lines) overlaid with the
    phonon coupling strength  (thick red regions):
    the region size is proportional to the corresponding $P_{\nu \bf q}$ value. 
    The vertical grey bands indicate sections of the first Brillouin zone
    where the resonance conditions are satisfied for $\hbar\omega =
    200$~meV with the chemical potential $\mu$ set to $-450$~meV and
    the energy zero corresponds to the Dirac point.  }
\end{figure*}

\subsection*{Pristine and disordered SLG samples}
Our samples consist of single layer graphene grown on copper foil using a
specific intermittent growth chemical vapor deposition (CVD) process that
allows production of a fully homogeneous monolayer without multilayer
patches~\cite{Han2014}.
The graphene thus obtained was then transferred to a MIR transparent
substrate with regular transfer techniques~\cite{Suk2011}.
The graphene sheets have been fully characterized through optical and scanning
electron microscopy, along with Raman microspectroscopy and hyperspectral
cartography to ensure that there is no bilayer contamination within the
investigated regions (see Supplementary Information).  
Covalent functionalization is then used to introduce disorder in the
crystalline lattice~\cite{Baranton2005,Gan2012a,Wang2012a}.
Indeed, grafting of iodophenyl moieties (\iph) to graphene breaks the \emph{sp$^2$}
conjugation and generates \emph{sp$^3$} hybridized scattering
centers~\cite{Niyogi2011}.
The reaction, however, leaves the samples in a state where the Fermi level is
poorly defined from adsorbed ions and grafts. 
We thus use chemical doping to set the chemical potential. 
Practically, this is
achieved by soaking the sample in a doping solution at a certain
concentration, and then drying in a stream of nitrogen, which leaves the
adsorbed dopant molecules on the surface.
For instance, 2,3-dichloro-5,6-dicyano-1,4-benzoquinone (DDQ, an electron
acceptor) dissolved in acetonitrile is used for \textit{p}-type doping,
while hydrazine (an electron donor) in acetonitrile allows reduction of
graphene.

\subsection*{Phonon-induced transparencies in MIR response}
Infrared spectroscopy measurements are presented in
Fig.~\ref{fig:slg-data}.
The first set of data presented in (a) shows the results of microscopy
measurements in reflection mode of SLG transferred onto intrinsic silicon
with 300~nm thermal oxide, which allowed us to probe well characterized
regions of the sample and to exclude areas with bilayer and few-layer
islands.
Except for a weak modulation near 1250~\wn\ and the carbonyl band of an impurity at
$\sim\,$1720~\wn\ (denoted by $\ast\ast$),
pristine graphene yields a featureless MIR spectrum in both
doping states (\textit{i.e.} [DDQ]$=1$~mM and [hydrazine]$=10$~mM)
[Fig.~\ref{fig:slg-data}(a), top panel]. 
This behavior is expected since no graphene phonons are supposed to be
infrared-active~\cite{Reich2004}.
In accordance with our following arguments, the weak modulation near
1250~\wn\ may be due to the fact that the probed area contains wrinkles 
and joint boundaries between crystal domains
(see microscopy pictures in Supplementary Information), which are
sources of disorder.
The observations, however, change drastically upon covalent grafting to
graphene [graft density of approximately 1~\iph\ for 100~C atoms as
evaluated by iodine content analysis by X-ray photoelectron spectroscopy
(XPS), see Supplementary Information]. 
As seen in the bottom panel of Fig.~\ref{fig:slg-data}(a), 
a broad asymmetric signal is now apparent near 1600~\wn\ in the
DDQ-doped trace,
and a second, even broader signal is also observed around 1250~\wn. 
Vibrational bands from dopants and \iph\ 
are absent from the spectra because their
concentrations are lower than the detection limit.
Moreover, the asymmetric bands disappear upon soaking in a hydrazine solution, and
can be recovered by doping back to \emph{p}-type (see Supplementary
Information), thus demonstrating a dependency on charge carrier density. 

The optical window of SiO$_2$ being limited in the infrared by an intense
absorption band at $\sim$1050~\wn, transmission spectra 
were obtained on a BaF$_2$ substrate to better assess the graphene band observed at
$\sim$1250~\wn.
In these measurements, macroscopic sample areas were probed to prove that
the effect is not limited to localized microscopic regions.
This second set of data is shown in Fig.~\ref{fig:slg-data}(b) using an
optical conductivity scale for
pristine graphene doped at [DDQ]$=1$~mM, and for grafted graphene (graft
density of approximately 2~\iph\ for 100~C atoms as evaluated by XPS)
\emph{p}-doped at
various levels (from [DDQ]$=1$~$\mu$M to 1~mM).
Raw spectra are presented in the top panel, while a cubic baseline has been
subtracted from all spectra in the lower panel.
Two antiresonances are identified at $\sim$1250~\wn\ and
$\sim$1600~\wn, whereas the pristine signal (black trace) is smooth, to the
exception of a spurious band denoted by a star ($\ast$), and the
carbonyl band of an impurity marked by double stars ($\ast\ast$).
The frequencies of the antiresonances coincide with those of the Raman $D$ and
$G$ bands, thus strongly suggesting that phonon modes are involved in the
 mechanism.
The profiles of the antiresonances are asymmetric and skewed toward higher energies.
The widths of the bands, as evaluated by fitting a Fano profile, are
approximatively  120~\wn\ and 45~\wn\ respectively for the lower and higher 
energy bands.
Moreover, the amplitude of these anomalies can be modulated with dopant
concentration, which becomes evident after baseline correction (bottom
panel).
As can be observed in the top panel, the background absorption also evolves
following the known dependence of the Drude peak to
doping~\cite{Horng2011}, thus supporting the fact that the signal cross
section is related to charge carrier density. 
It should be stressed that the observed transparency windows
(antiresonances) are important
hints about the underlying mechanism, since it is in discrepancy with normal phonon
resonances which absorb light and thus would yield upward bands on a conductivity scale. 

\subsection*{Extension to the Drude model}
To first gain intuition about the physics underpinning the
occurence of the antiresonances, we begin by extending
the Drude model to collisions between a representative semiclassical electron and 
a phonon mode. 
Requiring only momentum and energy conservation, we derive the following
expression for the conductivity $\sigma$ (Sec.~III-A,
Supplementary Information):
\begin{align}
    \label{eq:Delta Sigma}
    \mbox{Re}\big[\sigma(\omega)\big] &\simeq  \sigma_0\frac{(2\Gamma)^2}{(2\Gamma)^2+(\hbar\omega )^2}+\Delta \sigma(\omega)\\
     \frac{\Delta\sigma(\omega)}{\sigma_0} &\propto \frac{m}{M}
        \mbox{Im}\Bigg[\frac{(2\Gamma)^3}{(\hbar\omega +2i\Gamma)^2} \mathcal{D}_{ph}(\omega)\Bigg],
\end{align}
where $\sigma_0$ is the static Drude conductivity,  $m/M$ 
is the ratio of electronic to ionic mass, $2\Gamma$ is the width of the Drude peak
and $\mathcal{D}_{ph}(\omega)$ is the Green function of the phonon mode which
is sharply peaked at the frequency of the phonon.  The first term in Eq.~(\ref{eq:Delta Sigma}) is
the familiar Drude conductivity,
while the second term accounts for an interference effect induced by multiple 
subsequent electron-phonon scattering events and the dynamical nature of the phonon.
For $\hbar\omega > 2\Gamma$, this simple model reproduces
asymmetric antiresonances skewed toward higher energies and centered at the
phonon frequency (Fig.~S10, Supplementary Information).

\subsection*{Quantum mechanics model}
While this simple extension to the Drude model gives an intuitive picture for the 
underlying physics at play, it is not sufficiently sophisticated as it does not account for the 
quantum mechanical nature of electrons and phonons, nor does it take into
consideration selection rules or band dispersion.
A proper treatment of the optical conductivity makes use of the Kubo formula~\cite{Mahan}, which relates $\sigma(\omega)$ to the
current-current correlation function. The effects of electron-phonon coupling as well as defect scattering
can then be computed systematically using the machinery of perturbation theory and Feynman diagrams, 
an approach which treats both \textit{intraband} and \textit{interband} electronic excitations on equal footing.
The fully quantum mechanical mechanism corresponding to the simple model of Eq.~(\ref{eq:Delta Sigma})
is described by the Feynman diagram of Fig.~\ref{fig:current current fano} in the case where \textit{intraband}
electronic excitations (\textit{ie}, excitations near the Fermi energy) dominate. 
This same Feynman diagram, but with an emphasis on \textit{interband} electronic excitations,
was first proposed by Cappelluti~\etal{}~\cite{Cappelluti2010,Cappelluti2012}  to explain
the Fano profile observed in pristine bilayer graphene~\cite{Kuzmenko2009,Tang2010} 
and few-layer graphene~\cite{Li2012},
where a discrete mode (a tangential phonon mode at $\Gamma$) couples to the continuum of
electron-hole excitations responsible for an optical resonance.  
This mechanism was also applied to plasmon-phonon coupling to account for similar observations in 
bilayer graphene nanoribbons~\cite{Yan2014}.
In pristine single layer graphene, however, a Fano resonance has never been observed, 
nor is it expected:  indeed, the contribution from the mechanism of Fig.~\ref{fig:current current fano} vanishes by symmetry.
Even more, the other observed  band at $\sim$1250~\wn\ cannot be
attributed to coupling to $\Gamma$ point phonons, as there are no available
phonon modes near this energy. 

A key point to correctly model the observed phenomenon is the
introduction of lattice defects (\iph\ grafing), which destroys the
periodicity of the system. Its simplest consequence is
a reduction of the electronic lifetime, leading to a broader Drude peak. 
However, the introduction of electronic damping in the mechanism of 
Fig.~\ref{fig:current current fano} still leads to an expression which vanishes 
by symmetry in single layer graphene;
the effect of disorder must thus be treated beyond electronic lifetime reduction.
Averaging the current-current correlation function over all possible graft configurations, 
the simplest disorder contribution coupling non-$\Gamma$ phonons to electronic excitations
is given by the Feynman diagram of Fig.~\ref{fig:current current impurity}. It is shown  
in the Supplementary Information to yield a contribution to the conductivity of the form
\begin{multline}
    \label{eq:fano sigma}
    \mbox{Re}\Big[\sigma_{imp}(\omega)\Big] \simeq \frac{e^2}{\hbar} \big(n_{imp}a_0^2\big)
            \sum_{\alpha = x,y}\sum_{L =A,B}\\
            \frac{1}{N}\sum_{\nu \bf q}A_{\nu \bf q}^{\alpha L}
            \frac{|Q_{\nu \bf q}^{\alpha L}|^2-1 + 2 z_{\nu \bf q}\mbox{Re}\Big[ Q_{\nu \bf q}^{\alpha L}\Big]  }
            {|Q_{\nu \bf q}^{\alpha L}|^2\big(z_{\nu \bf q}^2+1\big)},
\end{multline}
with
\begin{align}
    z_{\nu \bf q} &\equiv \frac{\hbar\omega - \hbar\omega_{\nu \bf q}}{\Gamma_{\nu \bf q}},
\end{align}
where $N$ is the number of unit cells in the sample,
$n_{imp}$ is the number density of defects, $a_0$ is the Bohr unit of length, $-e$ the electronic
charge, $\alpha$ and 
$L$ are related to the spatial direction of the current operator ($x$ or $y$) and the defect scattering site in the graphene 
unit cell ($A$ or $B$). Also,  
$\nu$ is a phonon mode label and $\bf q$ is a momentum restricted to the First Brillouin zone; thus $\omega_{\nu \bf q}$ is the 
frequency and $\Gamma_{\nu \bf q}$ is related to the lifetime of the phonon labelled by  $(\nu \bf q)$. 
The contribution to the conductivity of Eq.~(\ref{eq:fano sigma}) is composed of a sum over the whole First Brillouin zone
of Fano-like terms with unitless Fano parameters $Q_{\nu \bf q}^{\alpha L}$ and amplitudes $A_{\nu \bf q}^{\alpha L}$.
These parameters are related to the ``Loop" function $\mathcal{H}^{\alpha L}_{\nu \bf q}$, corresponding to the 
sub-diagram of Fig.~\ref{fig:Hloop}, which can be expressed as the sum of a ``reactive" and ``absorptive" term,
\begin{align}
\mathcal{H}^{\alpha L}_{\nu \bf q} = \mathcal{R}^{\alpha L}_{\nu {\bf q}}  + i \mathcal{I}^{\alpha L}_{\nu {\bf q}},
\end{align}
such that, in atomic units,
\begin{align}
    Q_{\nu \bf q}^{\alpha L} &\equiv -\frac{\mathcal{R}^{\alpha L}_{\nu {\bf q}}}{\mathcal{I}^{\alpha L}_{\nu {\bf q}}} \quad\mbox{and}\quad
    A_{\nu \bf q}^{\alpha L} \equiv  
                \frac{1}{\hbar\omega_{\nu \bf q}} 
                         \frac{1}{4\Gamma_{\nu \bf q}}\Big| \mathcal{R}^{\alpha L}_{\nu {\bf q}}\Big|^2.
\end{align}
Given the asymmetrical nature of the spectrum,
it will be useful below to define the phonon coupling strength~\cite{Cappelluti2012},
\begin{align}
    \label{eq: phonon coupling strength}
    P_{\nu \bf q} &\equiv  
                \frac{1}{4}\sum_{\alpha=x,y} \sum_{L=A,B}\frac{1}{\hbar\omega_{\nu \bf q}} 
    \Big(\Big| \mathcal{R}^{\alpha L}_{\nu {\bf q}}\Big|^2 + \Big| \mathcal{I}^{\alpha L}_{\nu {\bf q}}\Big|^2\Big),
\end{align}
as a measure of the weight of the mode.

It is useful at this point to draw parallels between the proposed contribution to the current-current
correlation function and the theory of Raman scattering with a single phonon emission, presented diagrammatically 
in Fig.~\ref{fig:Raman}, along with the Loop function in Fig.~\ref{fig:Hloop}.
As depicted schematically in Fig.~\ref{fig:resonance condition}, the Loop function may
be described as the absorption of a photon to a virtual
excited state, followed by an elastic collision with a defect back to the
electronic band of graphene. 
If the whole scattering process is in resonance both in momentum and energy
with an optical phonon mode, then the cycle is similar to a Raman resonant 
transition, as also represented on Fig.~\ref{fig:resonance condition}, but with two major differences:
on the one hand, elastic scattering with defects allow for intra- and intervalley processes within the Loop function; 
large $\bf q$ phonon modes thus become available through this scheme.
Also, in Raman spectroscopy the incoming light is energetic enough to induce interband transitions, whereas infrared
light cannot: thus the resonant transition in the Loop function imposes that both momenta $\bf k$ and $\bf k+q$ 
be within $\hbar\omega$ of the Fermi energy, where $\hbar\omega$ is the energy of the incoming infrared light. 
There is always a $\bf k$ satisfying the resonant condition for 
\begin{align}
    \label{eq:condition}
    \frac{\omega}{v_F} \leq |{\bf q-P}| \leq 2 k_\mu +\frac{\omega}{v_F}
\end{align}
where $k_{\mu}$ is the radius of the circular Fermi surface and $v_F$ is the Fermi velocity of graphene.  
In the above, ${\bf P} = \Gamma$ for intravalley $\bf q$  and ${\bf P} = \pm$~K for intervalley $\bf q$.
If $\bf q$ is such that Eq.~(\ref{eq:condition}) cannot be satisfied, we expect 
nonresonant (and thus small) contributions to the current-current correlation function.

\subsection*{Simulation of optical conductivity}
To test the quantitative validity of the model,
the contribution of Eq.~(\ref{eq:fano sigma}) was computed within the 
tight-binding approximation in the spirit of the work of Peres \etal{}~\cite{Peres2006}.
Impurity scattering is modeled in terms of on-site, randomly located
impenetrable potentials, treated in a way to make use of the Full Born Approximation~\cite{Peres2006}. 
The phonon frequencies and polarization vectors were obtained from a force constant model~\cite{Dubay2003};
the phonon dispersion was modified in the vicinity of ~$\pm$K to account for the physically well established 
Kohn anomalies at these points~\cite{Lazzeri2008}, but which the force constant model fails to capture.
The parameters entering the electron-phonon coupling matrix elements
were obtained by comparing with computed results available in the literature
at ${\bf q}=\Gamma$ and ${\bf q} =$~K~\cite{Lazzeri2008,Jiang2005}
(see Supplementary Information).
The model contributions were computed for various chemical
potential values $\mu$ at a graft density of 2\%, a phonon energy broadening
$\Gamma_{\nu \bf q} = 2.5$~meV, and an electronic energy broadening
energy $\Gamma = 75$~meV (as exposed in Sec.~III-B of the
Supplementary Information, we estimate $\mu \simeq -450$~meV).

The model conductivity of Eq.~(\ref{eq:fano sigma})
is compared to the experimental conductivity in the presence of disorder and soaked in 
a 1~mM DDQ solution in Fig.~\ref{fig:results}.
The model reproduces the position and asymmetry of the two prominent bands
in the baseline-corrected signal, and yields a roughly correct
amplitude, given the uncertainty on the experimental chemical potential and
Drude peak width.
Also, it correctly captures the \emph{transmission windows} signal profile 
(\textit{i.e.} downward bands in the optical conductivity, upward bands on a
transmittance scale) due to an interference effect between electron
and phonon degrees of freedom, akin to the one captured by the intuitive extension of the Drude model.
Finally, the contribution of the model to the conductivity vanishes as
$\mu\rightarrow 0$, in agreement with the experimental observation that the
asymmetric bands disappear upon soaking in hydrazine solution.
For completeness, the conductivity was also computed for various
reasonable values of the broadening parameter $\Gamma$:
the asymmetry of the bands follows the same
trend as that of the simple Drude-like model of
Eq.~(\ref{eq:Delta Sigma}) even when disorder is treated beyond lifetime effects (See Fig.~S10 and ~S11(b)
 of the Supplementary Information).

Fig.~\ref{fig:phonon dispersion} shows that the phonon coupling strength
(red areas) is substantial only near regions where 
the resonance conditions are satisfied (grey bands).
This behavior leads to a coupling strength density of states with two
major peaks corresponding to contributions coming from ${\bf q} \simeq
\pm$~K and ${\bf q} \simeq \Gamma$, as plotted in Fig.~\ref{fig:phonon
strength} (top panel, right axis).

\section*{Discussion}
The two prominent antiresonances in the experimental data can now readily be explained. 
The feature at $\sim$1600~\wn\ comes from coupling to phonons near the $\Etwog$ mode for ${\bf q} \sim \Gamma$;
in agreement with experiments, the amplitude of the simulated band
increases with $|\mu|$, while the position of the band 
remains unchanged, reflecting the lack of phonon dispersion near the $\Etwog$ mode at $\Gamma$.
The broad feature in the vicinity of 1200-1400~\wn\ comes from coupling to phonons near the $\Aoneprime$
mode for ${\bf q} \sim\pm$~K; 
again, the simulated amplitude of the band increases with $|\mu|$ as observed
experimentally,
and its position moves to higher energy with $|\mu|$ [dashed line in
Fig.~\ref{fig:phonon strength} tracking the  position of the minimum], a
consequence of the strong phonon dispersion near the Kohn anomaly at
$\pm$~K. 
We note the absence of features around $\sim$1050~\wn, the frequency of 
the $\Atwoprime$ mode  at $\pm$~K: this absence is consistent with
simulations based on density functional theory suggesting that the electron-phonon coupling strength
of this mode is negligible~\cite{Piscanec2004}. 

Our experiments bring to light a disorder and phonon mediated phenomenon
in single layer graphene yielding transmission windows at phonon frequencies
matching ${\bf q}\sim \Gamma$ and ${\bf q}\sim\pm$~K.
Unlike the contribution leading to Fano resonances in pristine
multilayer graphene, the proposed mechanism is \emph{intraband} in nature and
accounts for disorder beyond lifetime effects:
electrons in states near the Fermi energy scatter coherently on
defects and phonons, thus breaking the optical selection rule
(\emph{i.e.} ${\bf q\simeq}\Gamma$) valid in pristine samples.
Whereas Fano profiles are induced by a \emph{discrete state}-\emph{continuum}
coupling between $\Gamma$ phonon modes and electronic degrees of freedom,
the mechanism presented here involves \emph{continuum}-\emph{continuum}
coupling, with phonon momenta constrained to small regions near $\Gamma$
and $\pm$~K.

This phenomenon may prove useful in quantifying the
disorder in graphene, and it will allow the
modulation of the optical conductivity in a narrow terahertz band,
hence providing extended tools for telecommunications, medical and security
imaging, and novel analytic and sensing capabilities.
Our model for single layer graphene is also expected to hold
in other systems: in
particular, the infrared spectroscopy of
bilayer~\cite{Kuzmenko2009,Tang2010} and few-layer~\cite{Li2012} graphene,
as well as carbon nanotubes~\cite{Lapointe2012}, should be revisited,
as there too shall disorder enable phonon-mediated intra- and intervalley scattering.

\bibliography{MIR_Graphene}

\section*{Acknowledgements}
The authors acknowledge insightful discussions with Didier Mayou and Gabriel Antonius, as well as technical
assistance from Pierre L\'evesque.

Research described in this paper was partly performed at the Mid-IR
beamline of the Canadian Light Source, which is supported by the Natural
Sciences and Engineering Research Council of Canada, the National Research
Council Canada, the Canadian Institutes of Health Research, the Province of
Saskatchewan, Western Economic Diversification Canada, and the University
of Saskatchewan.
The authors thankfully acknowledge funding from 
the Natural Sciences and Engineering Research Council of Canada
(NSERC) and le Fond de Recherche du Qu\'ebec - Nature et Technologies (FQRNT). 
Simulations were performed on the infrastructures of Calcul Qu\'ebec.


\section*{Authors Contributions}
B.R.\ and F.L.\ contributed equally to this article.
B.R.\ and M.C.\ devised the models and performed the simulations. 
F.L., E.G.\ and R.M.\ designed the experiments. 
Acquisition of the IR spectra was done by F.L., M.N., M.B.\ and R.M.
F.L.\ and M.N.\ functionalized graphene.
F.L., M.N., M.B., E.G.\ and S.C.\ proceeded to sample preparation and
characterization of the graphene layers.
Z.H.\ and V.B.\ provided the single layer graphene samples.
B.R., F.L., M.C.\ and R.M. wrote the manuscript.
All authors contributed to discussion and revision of the article.

\section*{Competing Financial Interests}
The authors declare no competing financial interests.

\end{document}